\documentclass[twocolumn]{aastex6}

\usepackage{apjfonts}
\usepackage{amsmath}
  
\newif\ifpdf\ifx\pdfoutput\undefined\pdffalse\else\pdfoutput=1\pdftrue\fi
\newcommand{\pdfgraphics}{\ifpdf\DeclareGraphicsExtensions{.pdf,.jpg}\fi}
\newcommand \microjy{$\mu$Jy}

\newcommand \spitzer{\textit{Spitzer}}
\newcommand \chandra{\textit{Chandra}}
\newcommand \xmm{\textit{XMM}}

\newcommand{\Lsun}{L_{\odot}}
\newcommand{\msigma}{$M_{\rm BH}-\sigma_{\rm bulge}$}

\def\gsim{\mathrel{\rlap{\lower4pt\hbox{\hskip1pt$\sim$}} \raise1pt\hbox{$>$}}}
\def\lsim{\mathrel{\rlap{\lower4pt\hbox{\hskip1pt$\sim$}} \raise1pt\hbox{$<$}}}

\slugcomment{Accepted for publication in the Astrophysical Journal;
  2017 Dec 4}
\shorttitle{}
\shortauthors{DONLEY ET~AL}

\begin{document}
\pdfgraphics

\title{Evidence for Merger-Driven Growth in Luminous, High-z, Obscured
  AGN in the CANDELS/COSMOS Field}

\author{J. L. Donley\altaffilmark{1},
J. Kartaltepe\altaffilmark{2},
D. Kocevski\altaffilmark{3},
M. Salvato\altaffilmark{4},
P. Santini\altaffilmark{5},
H. Suh\altaffilmark{6},
F. Civano\altaffilmark{7}, 
A. M. Koekemoer\altaffilmark{8}, 
J. Trump\altaffilmark{9},
M. Brusa\altaffilmark{10},
C. Cardamone\altaffilmark{11},
A. Castro\altaffilmark{12,13},
M. Cisternas\altaffilmark{14},
C. Conselice\altaffilmark{15},
D. Croton\altaffilmark{16},
N. Hathi\altaffilmark{17,8},
C. Liu\altaffilmark{18,19,20},
R. A. Lucas\altaffilmark{8}, 
P. Nair\altaffilmark{21}, 
D. Rosario\altaffilmark{4}, 
D. Sanders\altaffilmark{22}, 
B. Simmons\altaffilmark{23,24}, 
C. Villforth\altaffilmark{25,26},
D. M. Alexander\altaffilmark{27},
E. F. Bell\altaffilmark{28},
S. M. Faber\altaffilmark{29},
N. A. Grogin\altaffilmark{8},
J. Lotz\altaffilmark{8}, 
D. H. McIntosh\altaffilmark{30}, 
T. Nagao\altaffilmark{31}}

\altaffiltext{1}{Los Alamos National Laboratory, P.O. Box 1663, Los Alamos, NM 87545 jdonley@lanl.gov}
\altaffiltext{2}{Rochester Institute of Technology, Rochester, NY, USA}
\altaffiltext{3}{Colby College, Waterville, Maine 04901}
\altaffiltext{4}{Max Planck Institut f\"{u}r extraterrestrische Physik, Giessenbachstrasse 1, D-85748 Garching bei M\"{u}nchen, Germany}
\altaffiltext{5}{INAF-Osservatorio Astronomico di Roma, via di  Frascati 33, I-00040,Monte Porzio Catone, Roma, Italy}
\altaffiltext{6}{Subaru Telescope, National Astronomical Observatory of Japan, 650 A'ohoku place, Hilo, HI, 96720, USA}
\altaffiltext{7}{Harvard-Smithsonian Center for Astrophysics, 60 Garden Street, Cambridge, MA 02138, USA}
\altaffiltext{8}{Space Telescope Science Institute, 3700 San Martin Drive, Baltimore, MD 21218}
\altaffiltext{9}{University of Connecticut, 2152 Hillside Road, U-3046, Storrs, CT 06269}
\altaffiltext{10}{INAF-Osservatorio Astronomico di Bologna, via Ranzani 1, I-40127, Bologna, Italy} 
\altaffiltext{11}{Department of Math \& Science, Wheelock College,200 Riverway, Boston, MA 02215, USA}
\altaffiltext{12}{Universidad Nacional Aut{\'o}noma de M{\'e}xico (UNAM), Instituto de Astronom{\'i}a, Observatorio Astron{\'o}mico Nacional. A.P. 877, 22800 Ensenada, BC, M{\'e}xico}
\altaffiltext{13}{Universidad Aut{\'o}noma de Ciudad Ju{\'a}rez, Instituto de Ingenier{\'i}a y Tecnolog{\'i}a. 1210 Plutarco Elias Calles,32310 Cd. Juarez, CH, M{\'e}xico}
\altaffiltext{14}{Instituto de Astrofisica de Canarias, E-38205 La Laguna, Tenerife, Spain}
\altaffiltext{15}{School of Physics \& Astronomy, University of Nottingham, Nottingham, NG7 2RD, UK}
\altaffiltext{16}{Centre for Astrophysics \& Supercomputing, Swinburne University of Technology, P.O. Box 218, Hawthorn, Victoria 3122, Australia}
\altaffiltext{17}{Aix Marseille Universit\'{e}, CNRS, LAM (Laboratoire d'Astrophysique de Marseille) UMR 7326, 13388, Marseille, France}
\altaffiltext{18}{Astrophysical Observatory, Department of Engineering Science and Physics, City University of New York, College of Staten Island, 2800 Victory Boulevard, Staten Island, NY 10314}
\altaffiltext{19}{Department of Astrophysics and Hayden Planetarium, American Museum of Natural History, New York, NY 10024}
\altaffiltext{20}{Physics Program, The Graduate Center, CUNY, New York, NY 10016}
\altaffiltext{21}{Department of Physics and Astronomy, University of Alabama, Box 870324, Tuscaloosa, AL 35487-0324, USA}
\altaffiltext{22}{Institute for Astronomy, University of Hawaii, 2680  Woodlawn Drive, Honolulu, HI 96822, USA}
\altaffiltext{23}{Center for Astrophysics and Space Sciences, University of California, San Diego, La Jolla, CA 92093, USA}
\altaffiltext{24}{Einstein Fellow}
\altaffiltext{25}{University of Bath, Department of Physics, Claverton Down, Bath, BA2 7AY, UK}
\altaffiltext{26}{Scottish University Physics Alliance (SUPA), University of St Andrews, North Haugh, KY16 9SS, St Andrews, UK}
\altaffiltext{27}{Centre for Extragalactic Astronomy, Department of 
  Physics, Durham University, South Road, Durham DH1 3LE, UK}
\altaffiltext{28}{Department of Astronomy, University of Michigan, Ann Arbor, MI 48109, USA}
\altaffiltext{29}{Department of Astronomy and Astrophysics, University of California Observatories/Lick Observatory, University of California, Santa Cruz, CA 95064, USA}
\altaffiltext{30}{Department of Physics and Astronomy, University of Missouri-Kansas City, Kansas City, MO 64110, USA}
\altaffiltext{31}{Research Center for Space and Cosmic Evolution, Ehime University, Bunkyo-cho 2-5, Matsuyama, Ehime 790-8577, Japan}

\begin{abstract}

  While major mergers have long been proposed as a driver of both AGN
  activity and the \msigma\ relation, studies of moderate to high
  redshift Seyfert-luminosity AGN hosts have found little evidence for
  enhanced rates of interactions.  However, both theory and
  observation suggest that while these AGN may be fueled by stochastic
  accretion and secular processes, high-luminosity, high-redshift, and
  heavily obscured AGN are the AGN most likely to be merger-driven.
  To better sample this population of AGN, we turn to infrared
  selection in the CANDELS/COSMOS field.  Compared to their
    lower-luminosity and less obscured X-ray--only counterparts,
    IR-only AGN (luminous, heavily obscured AGN) are more likely to be
    classified as either irregular (50$^{+12}_{-12}$\%
    vs. 9$^{+5}_{-2}$\%) or asymmetric (69$^{+9}_{-13}$\%
    vs. 17$^{+6}_{-4}$\%) and are less likely to have a spheroidal
    component (31$^{+13}_{-9}$\% vs. 77$^{+4}_{-6}$\%).  Furthermore,
    IR-only AGN are also significantly more likely than X-ray--only
    AGN (75$^{+8}_{-13}$\% vs.  31$^{+6}_{-6}$\%) to be classified
    either as interacting or merging in a way that significantly
    disturbs the host galaxy or disturbed though not clearly
    interacting or merging, which potentially represents the late
    stages of a major merger.  This suggests that while major mergers
    may not contribute significantly to the fueling of Seyfert
    luminosity AGN, interactions appear to play a more dominant role
    in the triggering and fueling of high-luminosity heavily obscured
    AGN.

\end{abstract}


\section{Introduction}

The origin of the evolutionary connection between supermassive black
holes (SMBHs) and their host galaxies, as evidenced by the \msigma\
relation \citep{magorrian98,gebhardt00}, remains one of the major open
questions in extragalactic astronomy. For the past decade, theorists
have invoked major mergers between gas-rich galaxies to explain not
only this correlation, but also the strikingly similar cosmic
evolution of AGN and star-formation activity (e.g.,
\citealt{hopkins08}).

Not only are major mergers required to reproduce the properties of
classical bulges in simulations of galaxy formation, but when coupled
with feedback, mergers can reproduce many of the global properties of
both the AGN and galaxy populations, including the \msigma\ relation
\citep{dimatteo05, robertson06, hopkins06, hopkins08}. Furthermore,
the best examples of ongoing mergers in the local Universe,
ultraluminous infrared galaxies (ULIRGS), have long been known to be
active sites of both intense star-formation and AGN activity. This
observation led to a proposed co-evolutionary scenario in which major
mergers drive the growth of bulges via nuclear star-formation and
violent relaxation, and provide fuel to a rapidly accreting AGN via
merger-induced gravitational torques \citep{sanders88,hopkins08}.

At first glance, studies of local ($z < 0.45$) QSOs appear to support
this scenario, with 25-100\% showing evidence for ongoing mergers or
tidal debris \citep{bahcall97, canalizo01, zakamska06,bennert08,
  greene09, veilleux09b}. While these small targeted studies indicate
that local QSOs are commonly associated with mergers, the few studies
that compare their morphologies to inactive control samples fail to
find evidence for enhanced morphological disturbances \citep{dunlop03,
  reichard09}. QSOs, however, are rare in the local Universe, and may
not be triggered by the same mechanisms responsible for driving their
high-redshift counterparts onto the \msigma\ relation, which was
largely in place by $z \sim 1$ (e.g., \citealt{cisternas11a}).  To
determine the prevalence of mergers among high-redshift AGN, we have
therefore turned to the cosmological deep fields.

Studies of AGN hosts in the GEMS, GOODS, AEGIS, and COSMOS fields have
predominantly targeted X-ray selected AGN at $0.2 < z < 1.3$
\citep{sanchez04, grogin05,
  pierce07,gabor09,cisternas11b,simmons12,villforth14, rosario15,
  bruce16}. While a small fraction of these AGN hosts show strong
distortions, the rate of morphological disturbances is similar to that
of inactive galaxy control samples, suggesting that mergers do not
play a dominant role in AGN fueling, at least out to $z \sim 1$.  That
said, there is evidence for a factor of $\sim 2.5$ enhancement of
Seyfert-level AGN activity in close pairs
\citep{silverman11,ellison11} and evidence that minor mergers may play
a role in the fueling of moderate-luminosity AGN \citep{altamirano16}.
Furthermore, \cite{koss10} find both an increased pair and merger
fraction in local hard X-ray AGN samples.

To extend this analysis to the peak of AGN activity at $z \sim 2$, we
turn to the near-IR HST/WFC3 CANDELS survey, which probes light
redward of the 4000~\AA\ break and thus traces emission from the stars
responsible for the bulk of the stellar mass. Initial studies of the
hosts of $z \sim 2$ X-ray AGN in the GOODS-S region of CANDELS,
however, likewise find morphologies that are indistinguishable from
those of normal star-forming galaxies
\citep{schawinski11,kocevski12,simmons12}. Furthermore, the high
incidence of disk galaxies, which should be destroyed by major
mergers, suggests that a time delay between merger activity and the
AGN phase cannot account for the lack of merger signatures.

While these observations appear to call into question the role of
major mergers in AGN/galaxy co-evolution, most X-ray AGN populations
studied thus far are dominated by low-luminosity Seyfert galaxies
(log~$L_{\rm 0.5-10 keV}$~(ergs~s$^{-1})< 44)$, which may be
experiencing a different mode of SMBH and galaxy growth than their
high-luminosity counterparts. For instance, while cosmological
simulations require mergers to reproduce the properties of luminous
QSOs, stochastic accretion and secular processes can account for the
lower levels of nuclear activity in Seyfert galaxies
\citep{hopkinshernquist06,hopkins08,hasinger08, hopkins09, hopkins14}.
This hypothesis appears to be backed by a growing number of studies
that find a larger merger-driven and disturbed fraction among high
luminosity AGN across a range of redshifts \citep{guyon06, urrutia08,
  kartaltepe10, koss12, treister12, glikman15, ellison16, fan16},
though there are exceptions \citep{villforth14, mechtley16,
  villforth17}.  A similar trend has been observed in both the local
and high-z infrared and SMG galaxy populations, where late-stage major
mergers are responsible for fueling nearly all of the most luminous
galaxies \citep{larson16, engel10, ivison12}.

If the merger-driven evolutionary scenarios summarized in
\cite{hopkins08} and \cite{alexanderhickox12} are correct, the early
phases of a major merger should be dominated by luminous, yet heavily
obscured, AGN activity.  As the SMBH grows, AGN feedback then serves
to remove the dust and gas and the AGN becomes dust reddened and
eventually unobscured, but only after the fading merger features
become difficult to identify, particularly in the distant universe. To
test the major merger scenario for the co-evolution of SMBHs and their
hosts, studies should therefore target not only luminous AGN, but
heavily obscured luminous AGN.  While doing so can be difficult using
soft X-ray and optical emission alone, the same dust and gas that
serves to obscure the AGN's signatures also acts like a natural
coronagraph, blocking the intense UV-optical radiation from the AGN
itself and permitting a study of the underlying host galaxy emission.

A number of studies have begun to target obscured AGN at both low and
high luminosity, and most (\citealt{schawinski12} is an exception) have
indeed found a higher rate of disturbances among more heavily obscured
samples \citep{koss10,
  urrutia12,satyapal14,kocevski15,ellison16,shangguan16,fan16,weston17} or
evidence that extinction peaks during the intermediates stages of
merger evolution \citep{veilleux09}, albeit with a strong chaotic
component.  This suggests that AGN unification is not due solely to
orientation \citep{cattaneo05,kocevski15}.

High luminosity, high redshift, and heavily obscured AGN therefore
comprise the population of AGN most likely to experience merger-driven
SMBH and galaxy co-evolution.  Fortunately, these AGN can effectively
be targeted using their mid-IR colors.  For an AGN to be identified via
its MIR colors, its hot dust emission must overpower the underlying
stellar emission from the host galaxy (e.g.,
\citealt{donley08,donley12}). MIR selection therefore identifies the
most luminous, and thus the highest-redshift ($z \sim 2$) AGN in the
limited volumes probed by deep survey fields, recovering few Seyfert
galaxies but $>75\%$ of X-ray AGN with QSO luminosities
\citep{donley12}. Furthermore, because NIR-MIR emission is largely
insensitive to intervening obscuration, the unique MIR power-law
signature of luminous AGN is observable in both unobscured and heavily
obscured AGN.  MIR selection therefore provides a way to recover
highly complete samples of luminous, high-z, and heavily obscured AGN,
exactly those AGN whose evolution is expected to be dominated by major
mergers \citep{aah06,donley07,donley10,donley12}.

In the work that follows, we directly compare for the first time the
morphologies of X-ray and infrared selected AGN.  If mergers do indeed
play a dominant role in the triggering of luminous, high-z, obscured
AGN, we should see an excess of merger signatures in our sample of
high-luminosity, heavily obscured IR-only AGN when compared to the
lower-luminosity, less heavily-obscured X-ray--only population.

This paper is organized as follows.  In \S2 we discuss the relevant
data in the CANDELS/COSMOS field and describe our selection of the
infrared and X-ray AGN samples. The sample properties (reliability,
redshifts, luminosities, and stellar masses) are given in \S3. In
\S4 we present the visual classification scheme as well as the
resulting morphologies for the infrared and X-ray AGN samples.  The
conclusions are given in \S5.  Throughout the paper, we assume the
following cosmology: ($\Omega_{\rm m}$,$\Omega_{\rm
  \Lambda},H_0$)=(0.27, 0.73, 70.5~km~s$^{-1}$~Mpc$^{-1}$).

\section{Data and Sample Selection}

We selected the AGN for this study from the $\sim 200$ sq. arcmin
CANDELS/COSMOS field.  The CANDELS survey imaged the central region of
the COSMOS field in both the WFC3 F125W (J-band) and F160W (H-band)
filters to 5$\sigma$ limiting AB magnitudes of 27.72 and 27.56,
respectively (see \citealt{koekemoer11} and \citealt{nayyeri17} for details
on the CANDELS HST data products).  This high resolution NIR imaging
data provides a crucial look at the rest-frame optical emission from
moderate to high redshift galaxies and AGN.

The COSMOS survey provides ample multiwavelength data over the field,
including the X-ray (\xmm\ and \chandra) and MIR (\spitzer\ IRAC)
coverage most relevant to this work.  Shallow ($\sim 40$~ks) \xmm\
data extends over the full COSMOS field
\citep{hasinger07,cappelluti09,brusa10}, whereas the deeper \chandra\
coverage used here is limited to the central 0.9~deg$^2$, but fully
covers the CANDELS/COSMOS field with an average exposure time of $\sim
170$~ks.  (Deeper \chandra\ data now exist in the outer regions of the
COSMOS field \citep{civano16}, but do not overlap the CANDELS field.)

The primary \spitzer\ IRAC data used for this study \citep{sanders07}
cover the full COSMOS field to 1200s depth, or 5$\sigma$ sensitivities
of 0.9, 1.7. 11.3, and 14.6 \microjy\ in the 3.6, 4.5, 5.8, and 8.0
\micron\ bands, respectively.  We exclude from our study all IRAC
sources that lie within the masked regions of bright ($K<14$) 2MASS
sources, but include sources with flags indicating nearby neighbors or
deblending.  Of the $\sim 1000$ IRAC sources that fall within the
CANDELS/COSMOS field and that are brighter than the 5$\sigma$
sensitivities in each of the IRAC bands, 9\% have either bad pixels or
neighbors bright and close enough to significantly bias the photometry
and 11\% were originally blended with another object (note: these are
not mutually exclusive).  Of the 43 infrared-selected AGN (IRAGN) we
will select using this data, 5 (12\%) are flagged as having nearby
neighbors or were blended with another object, but a careful
inspection of each source indicates that the IRAGN selection is robust
to these issues.  Furthermore, our comparison below with the COSMOS15
IRAC photometry (see \S3.1) provides an additional check on the IRAGN
selection reliability.

\subsection{IRAGN} 

IRAGN were selected directly from the \cite{sanders07} IRAC catalog
using the criteria outlined in
\cite{donley12}: 

\begin{eqnarray}
& x = \rm{log_{10}}\left(\frac{f_{5.8\mu m}}{f_{3.6\mu m}}\right), \hspace*{0.2cm} y = \rm{log_{10}}\left(\frac{f_{8.0\mu m}}{f_{4.5\mu m}}\right) \\
 &\hspace*{0.1cm} x \ge 0.08 \hspace*{0.2cm} \wedge \hspace*{0.2cm} y \ge 0.15 \hspace*{0.2cm}  \\
\nonumber & \wedge \hspace*{0.2cm} y \ge (1.21\times x) - 0.27 \hspace*{0.2cm}\wedge \hspace*{0.2cm} y \le (1.21\times x) + 0.27  \\
\nonumber & \wedge \hspace*{0.2cm} f_{4.5\mu m} > f_{3.6\mu m} \hspace*{0.2cm}  \wedge \hspace*{0.2cm} f_{5.8\mu m} > f_{4.5\mu m} \hspace*{0.2cm} \wedge \hspace*{0.2cm} f_{8.0\mu m} > f_{5.8\mu m} 
\end{eqnarray}
\vspace{0.2cm}

\noindent As in \cite{donley12}, we require that IRAGN have fluxes
that exceed the $5\sigma$ sensitivities in each of the IRAC bands (see
above).  We experimented with loosening this criterion, but the vast
majority of additional sources we select were not clearly visible in
one of either the 5.8\micron\ or 8.0\micron\ bands.  In total, we
identify 43 IRAGN across the CANDELS/COSMOS field.  We cross-check the
IRAGN selected using the \cite{sanders07} catalog against those
selected using the \cite{laigle16} COSMOS15 catalog in \S3.1 below.

After selecting the IRAGN, we located the nearest CANDELS H-band
source using a search radius of 2\arcsec.  By directly matching the
IRAC and H-band catalogs, we avoided imposing a criterion that there
be a visible (I-band) counterpart.  In most cases, this made no
practical difference, but for 3 IRAGN, the nearest H-band source has
no corresponding optical counterpart in the Subaru I-band catalog of
\cite{ilbert09}.  In contrast, all IRAGN had an H-band counterpart,
with median and maximum offsets of 0.12\arcsec\ and 0.56\arcsec,
respectively.

\subsection{X-ray AGN}

X-ray AGN were selected from the \chandra\ catalog of
\cite{civano12}. \footnote{60 sources in our sample have \xmm\
  counterparts from \cite{brusa10}.  Of these, all but 6 have
  \chandra\ counterparts. We carefully examined these six sources, and
  found that only one, also an IRAGN, had a visible excess of
  \chandra\ counts near the \xmm\ source position.  While we therefore
  consider this source to be X-ray detected, we exclude the remaining
  \xmm-only sources from our sample as all are detected in only one of
  the \xmm\ bands (full/hard/soft) to $\le 6 \sigma$.  This choice has
  no effect on the final conclusions of this work.} We match the
H-band catalog directly to the X-ray positions.  Of the 99 \chandra\
sources, all but 3 have a clear H-band counterpart within the 2''
radius.  One of these sources is later removed from our sample because
it falls below our X-ray luminosity cut.  For the remaining two
sources, we searched for an H-band counterpart using the optical
counterpart position given in \cite{civano12}.  One has an H-band
counterpart only 0.11\arcsec\ from the optical position, the other has
no H-band counterpart, and is therefore excluded from our study.

Starting from the H-band, as opposed to the optical I-band, allowed us
to identify clear counterparts for 4 \chandra\ sources that have no
Subaru optical counterpart \citep{ilbert09} in \cite{civano12}.  For
the remaining sources, the optical counterpart nearest our H-band
counterpart matches the optical counterpart identified by
\cite{civano12}.

\section{Sample Properties}

Our full sample of X-ray and infrared-selected AGN consists of 43
IRAGN, 16 of which have no X-ray counterpart, and 72 Chandra or
XMM-selected AGN that are {\it not} IRAGN (or that fall below the IRAC
flux cuts in one or more of the IRAC bands).  IRAGN selection
predominantly identifies the most intrinsically luminous AGN (see
  \S3.4), and the IRAGN lacking X-ray counterparts are likely to be
luminous but heavily obscured \citep{donley12}.  In contrast, the
X-ray--only sample should predominantly lie at lower luminosities
and/or redshifts and be dominated by sources with low to moderate
obscurations.  As such, it will serve as our control sample for
comparison to the higher luminosity and higher-redshift IRAGN.

\subsection{IRAGN Selection Reliability}

To check the reliability of the IRAGN selected using the
\cite{sanders07} catalog, we turn to the COSMOS15 catalog
\citep{laigle16}, a NIR-based catalog with PSF-matched photometry from
the UV to the MIR.  COSMOS15 takes advantage of the deeper data in
IRAC channels 1 and 2 provided by the SEDS and SPLASH COSMOS surveys
(see Capak et al. 2017, in prep), but uses the same data presented in
\cite{sanders07} for IRAC channels 3 and 4.  However, for these two
longer-wavelength channels, the error estimates from COSMOS15 are far
more conservative than those given by \cite{sanders07}, and the
agreement between the catalogs begins to break down for sources with
moderate \cite{sanders07} $S/N$ but COSMOS15 $S/N < 3$ in one or more
of the IRAC bands.

If we require that our IRAGN be selected as such based on both the
\cite{sanders07} and COSMOS15 catalogs, our sample of X-ray--detected
IRAGN drops by four from 27 to 23.  The four sources that are lost all
have COSMOS15 $S/N > 3$ in each of the IRAC bands, but the modest
differences in photometry tend to place the \cite{sanders07}-selected
IRAGN just outside of the selection box.  

Because X-ray undetected IRAGN tend to be fainter than their
X-ray--detected counterparts, the discrepancy between the catalogs at
low COSMOS15 $S/N$ has a far larger effect on our sample of X-ray
undetected IRAGN.  Of the 16 \cite{sanders07}-selected IRAGN without
X-ray counterparts, only 7 would also be selected as IRAGN using the
COSMOS15 photometry (all 7 have COSMOS15 $S/N$ > 3 in each IRAC
band).  We make the case below for keeping an additional 2 sources,
bringing the number of X-ray undetected AGN to 9, but the cross-check
with COSMOS15 may remove as many as 7 of the 16 IRAGN identified by
\cite{sanders07}.

Of these 7 IRAGN not selected using COSMOS15, 5 have a COSMOS15 $S/N
\lsim 3$ in IRAC channel 4 and largely discrepant channel 4 fluxes
between the two catalogs, and 2 have $S/N > 3$ but were already
marginal IRAGN.  Whether these sources are indeed IRAGN or not
therefore appears to be catalog-dependent, and we will consider both
cases in the analysis below.  As for the remaining two
X-ray--undetected IRAGN not identified by COSMOS15, one lacks a
COSMOS15 counterpart altogether but has a high \cite{sanders07} $S/N$
in all IRAC bands and the other is a single IRAC source whose flux
appears to have been split between two optical/NIR counterparts in
COSMOS15.  Furthermore, for the latter, only its slightly
non-monotomic COSMOS15 IRAC fluxes cause it to not meet our strict
IRAGN criteria: its COSMOS15 photometry places it inside the IRAGN
selection box.  We therefore keep both IRAGN in our COSMOS15 sample,
which can generally be viewed as a higher $S/N$ subsample of the full
\cite{sanders07}--selected IRAGN sample.  The impact of this IRAC
$S/N$ cut on our results will be discussed below in \S4.1.

\subsection{Redshifts}

Of the 115 AGN in our full sample, 69 have spectroscopic redshifts
from either public or internal COSMOS team datasets obtained using the
following instruments or surveys: DEIMOS (Keck), FMOS (Subaru), FORS1
(VLT), FORS2 (VLT), FOCAS, HST Grism, IMACS (Magellan), LRIS (Keck),
MOSFIRE (Keck), PRIMUS (Magellan/IMACS), SDSS, VIMOS (zCOSMOS),
XSHOOTER (VLT), and 3DHST.  The spectroscopic redshift fraction is
$\sim 60\%$ for both the IRAGN and X-ray--only samples.  For the
remaining X-ray detected AGN with optical counterparts, we adopt the
AGN-specific photometric redshifts calculated by \cite{salvato11}, and
for the IRAGN lacking X-ray counterparts, we calculate photometric
redshifts using the methods outlined in \cite{salvato11} for
consistency.  Doing so gives redshift measurements or estimates for
all but the 3 IRAGN and 4 X-ray--only AGN that lack clear optical
counterparts.  The median redshifts for the \cite{sanders07} and
COSMOS15 IRAGN samples are $z=1.93$ and $z=1.87$, respectively, and
that of the X-ray--only sample is $z=1.22$.

\subsection{Observed X-ray Luminosities}
We plot in Figure 1 the redshifts and observed 0.5-10~keV X-ray
luminosities for our AGN samples.  We have excluded from the
X-ray--only sample 4 X-ray sources with luminosities lower than
$10^{42}$~erg~s$^{-1}$.  We also identify in Figure 1 those AGN that
meet the IRAGN criteria.

As can be seen in Figure 1, 80\% of the high luminosity ($L_x >
10^{44}$~erg~s$^{-1}$) X-ray AGN with good IRAC fluxes are also
IRAGN. In contrast, only 25\% of the lower-luminosity ($L_x <
10^{44}$~erg~s$^{-1}$) X-ray AGN with good IRAC fluxes are IRAGN, and
80\% of these have X-ray luminosities greater than
$5\times10^{43}$~erg~s$^{-1}$.  As expected, the IRAGN selection
effectively identifies the highest luminosity AGN, whereas the X-ray
selection is sensitive to lower-luminosity Seyfert galaxies.

\begin{figure}[t]
\includegraphics[width=\linewidth]{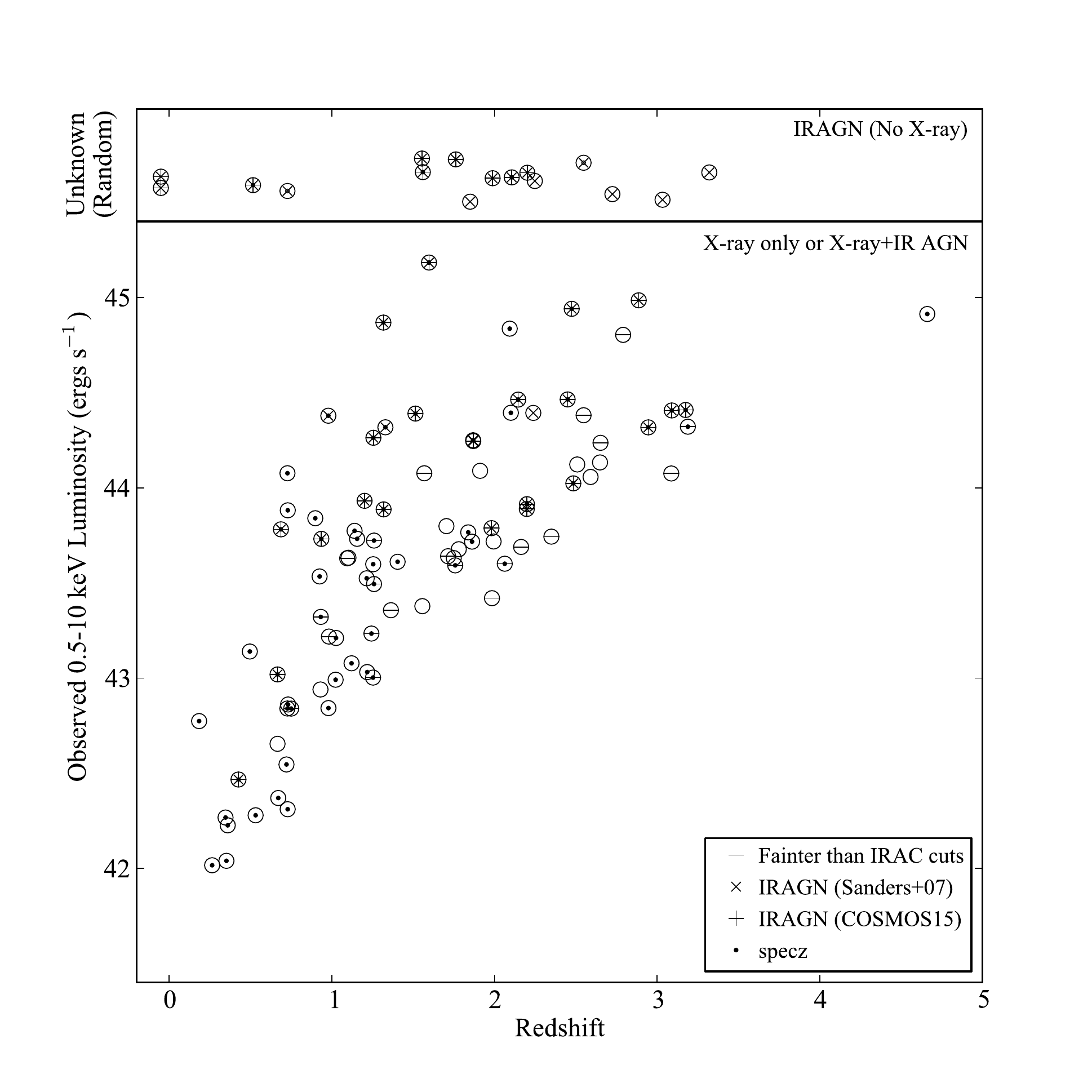}
\caption{Observed 0.5-10~keV X-ray luminosity vs. redshift for the IR
  and X-ray selected AGN.  IR-only AGN are shown on the top, where
  their unknown ``X-ray luminosity'' has been randomized for plotting
  purposes and sources with unknown redshifts are assigned a value of
  -0.1.  The vast majority of high-luminosity X-ray AGN are also
  selected as IRAGN, whereas most X-ray--only AGN are Seyfert
  galaxies. }
\end{figure}

\begin{figure*}[t!]
$\begin{array}{cc}
\includegraphics[width=0.5\linewidth]{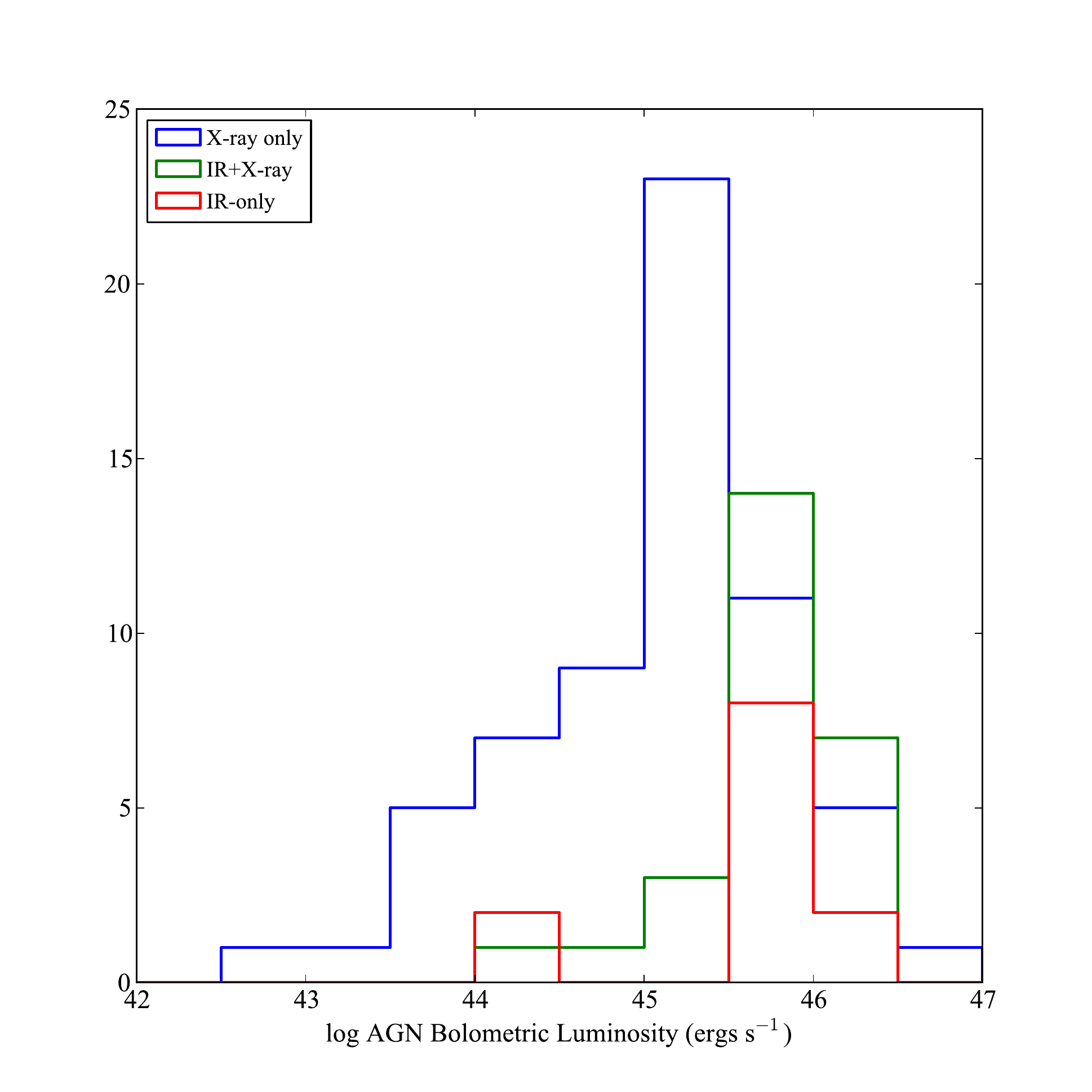} 
\includegraphics[width=0.5\linewidth]{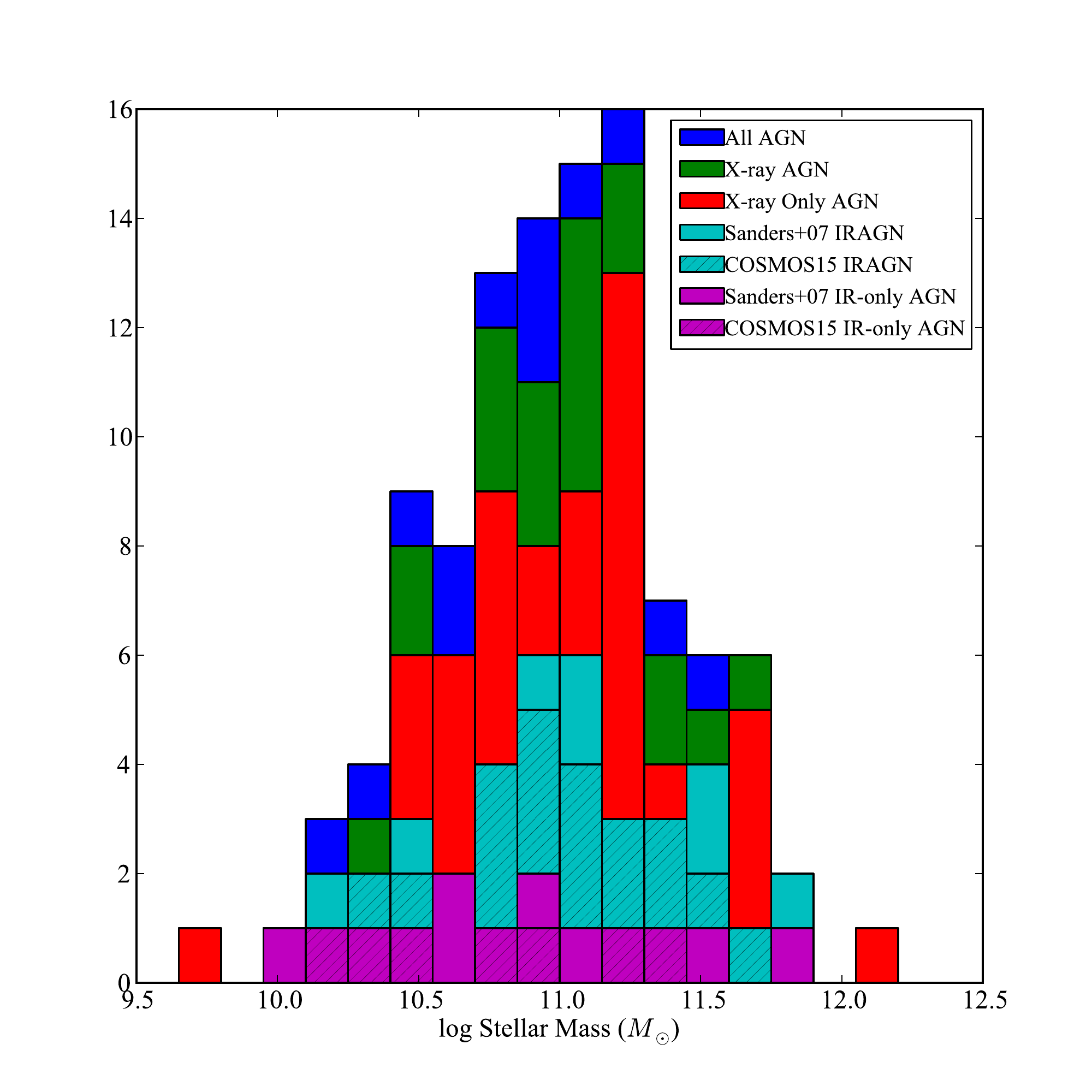} 
\end{array}$
\caption{Left: AGN bolometric luminosity for the IR-only, IR+X-ray,
  and X-ray--only AGN populations.  Bolometric luminosities were
  derived from SED fitting for sources with 24\micron\ counterparts,
  and from the X-ray luminosity for sources without 24\micron\
  counterparts. IRAGN selection preferentially identifies the most
  intrinsically luminous AGN.  Right: stellar mass distribution for
  the various AGN subsamples, which are consistent with having been
  drawn from the same population.}
\end{figure*}

\subsection{Bolometric Luminosities}

Using the COSMOS15 catalog and the SED-fitting approach of
\cite{suh17}, which requires both a redshift and a 24\micron\
detection, we calculate the AGN bolometric luminosity for 76 of the
111 AGN in our sample.  The requirements listed above exclude 4
IR-only AGN (one with no COSMOS15 counterpart, one with no redshift,
and two that are blended with brighter nearby sources and so have no
reported 24\micron\ fluxes), as well as 31 X-ray sources (2 with no
COSMOS15 counterpart, 4 with no redshift, and 25 with no 24\micron\
counterpart).

For X-ray AGN with redshifts, we can also estimate the AGN bolometric
luminosity using the absorption-corrected rest-frame 0.5-10~keV X-ray
luminosities from \cite{marchesi16a} and \cite{marchesi16b}, where we
give preference to the results from X-ray spectral fitting
\citep{marchesi16b} when they are available\footnote{We apply a
  correction factor to the X-ray luminosities from \cite{marchesi16a}
  to correct for a systematic offset between this catalog and
  \cite{marchesi16b} due to an inconsistency between the assumed
  spectral shapes (F. Civano 2017, private communication).}  (Note:
for sources with only an upper limit on N$_{\rm H}$, we apply no
absorption correction, and for sources with only a lower limit on
N$_{\rm H}$, we apply the correction associated with this lower
limit.) Of the 94 \chandra\ sources in our sample, 62 have both an
SED-derived bolometric luminosity as well as an X-ray
luminosity. Comparing the AGN luminosities for this subsample gives an
X-ray bolometric correction of k$_{\rm bol} = 44$, in agreement with
\cite{hopkins07} for an AGN sample with the median bolometric
luminosity of our sample: 11.9~$\Lsun$.  The scatter about the
resulting L$_{\rm bol}$(SED) vs. L$_{\rm bol}$(X-ray) relation has a
standard deviation of $\sigma$~(log L$_{\rm bol}$) = 0.45.  Using this
bolometric correction, we estimate the AGN bolometric luminosity for
the remaining 26 X-ray sources in our sample with a redshift but no
24$\micron$ counterpart.  Combining these X-ray--derived AGN
bolometric luminosities with the SED-derived AGN bolometric
luminosities gives the distributions shown in Figure 2, where we give
preference to the SED-derived L$_{\rm bol}$ when it is available.

As expected, and as was demonstrated using the X-ray luminosities in
Figure 1 for AGN with X-ray counterparts, IRAGN selection
preferentially identifies the most intrinsically luminous AGN.  (The
two IR-only AGN with log~L$_{\rm bol}$(ergs s$^{-1}) <44.5$ are the
two IRAGN in Figure 1 with $z<1$.) 

\subsection{Stellar Masses}

To ensure that our morphological analysis is not biased by the
underlying stellar mass distribution of the various AGN samples, we
also plot in Figure 2 the distribution of stellar masses calculated
using the techniques described in \cite{santini12}, which take into
account both the stellar and AGN contributions to the SED.  As can be
seen, there is no systematic offset between the stellar masses of the
X-ray and IR-selected samples.  Instead, as confirmed by KS tests
between the full AGN sample and various subsamples, as well as between
the IR-only and X-ray--only subsamples (p-value~$=0.46$), the AGN
sub-samples are well-matched in stellar mass.

\section{Visual Morphologies}

To determine the morphologies of our AGN sample, we utilized the
CANDELS visual classification framework as described in
\cite{kartaltepe15}.  For this study, the classification GUI displayed
the Subaru V-band, HST ACS F814W (I-band), HST WFC3 F125W (J-band),
and HST WFC3 F160W (H-band) images for each AGN. Twenty-one
classifiers (all professional astronomers) then chose one or more of
the following morphology classes for each AGN: disk, spheroid,
peculiar/irregular, compact/point source, and unclassifiable.  They
then selected just one of the following interaction classes: merger,
interaction within the segmentation map, interaction outside the
segmentation map, non-interacting companion, and no interaction.
Classifiers also had the option of selecting from a number of flags,
including tidal arms, double nuclei, asymmetric, and point source
contamination.  It is worth noting that while classifiers knew they
were classifying a sample of AGN, they did not know which AGN were
X-ray or IR-selected.

From these raw classifications, we created the following consensus
classification categories for each source.  For morphology, we added
the asymmetric flag to our morphology classification (and flagged
asymmetric in all cases where irregular had been chosen), resulting in
the following non-mutually exclusive classes: disk, spheroid,
irregular, point source (which also includes the point-source
contamination flag), asymmetric, and unclassifiable.  Following
\cite{kocevski12}, our consensus morphology class was then taken to be
the combination of classes chosen by at least half (11+) of the
classifiers.  We show an example in Figure 3, where we plot the
thumbnail images and individual morphology classifications for one of
the IRAGN not detected in X-rays.  In this particular case, the
consensus morphology is Irregular+Asymmetric.

To determine the interaction class, we created the following five
categories: undisturbed (no companion, interaction, merger, asymmetry,
double nuclei, or tidal arms), undisturbed with a companion
(companion, but no interaction, merger, asymmetry, double nuclei, or
tidal arms), disturbed (no clear interaction or merger, but yes to
asymmetry, double nuclei, or tidal arms), interaction or merger
(either inside or outside the segmentation map), and unclassifiable.
Our consensus interaction class was taken to be the most commonly
selected of these five mutually-exclusive classes.  In the example
shown in Figure 3, nearly all of the classifiers agree that this
particular AGN is in an interacting or merging system.

\begin{figure}[t]
\includegraphics[width=\linewidth]{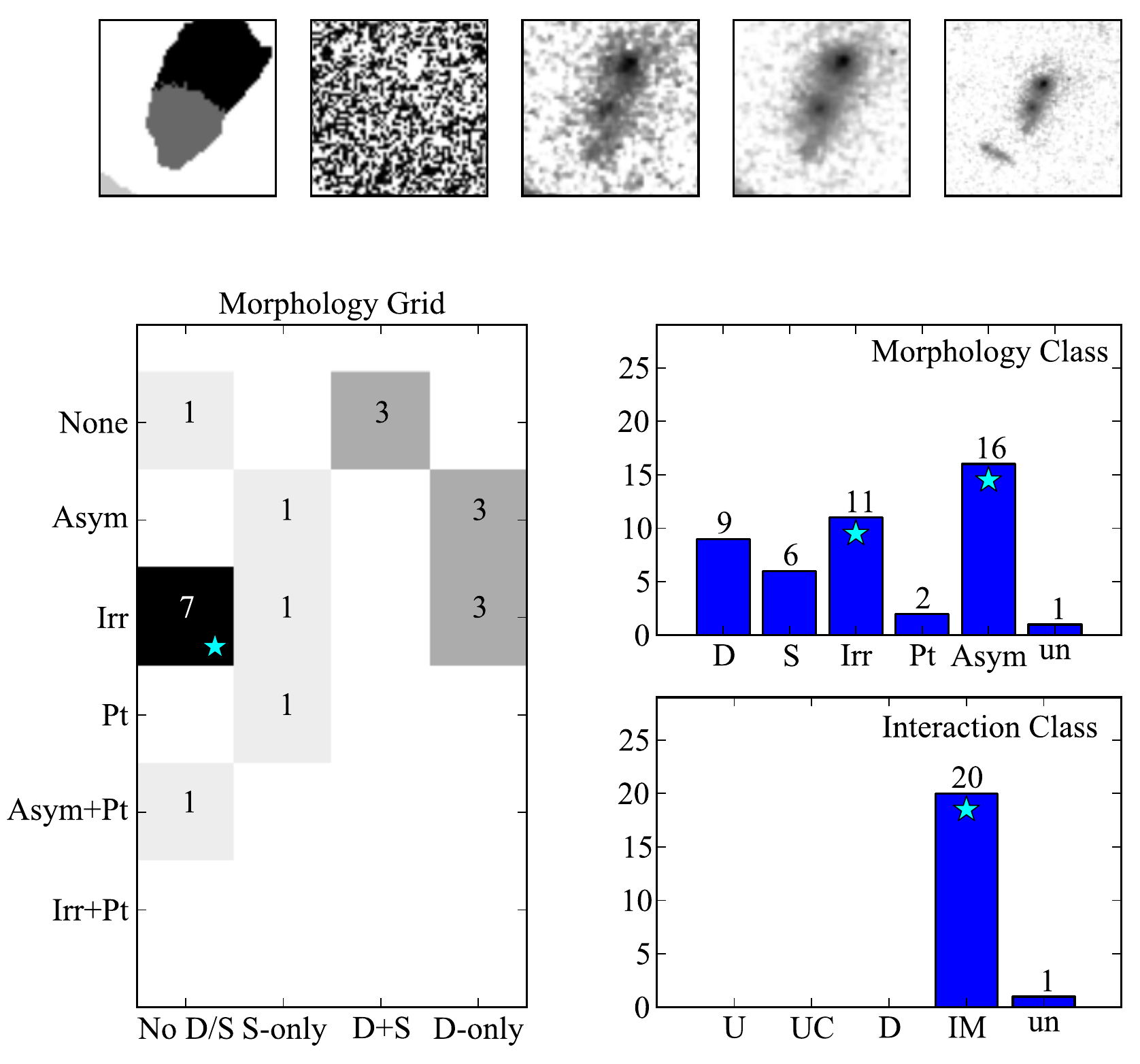}
\caption{Morphology classification for one of the X-ray non-detected
  IRAGN.  From left to right, the thumbnails at the top show the
  segmentation map along with the ACS F814W I-band image, WFC3 F125W
  J-band image, and WFC3 F160W H-band images both at the nominal size
  and twice that to identify nearby companions. The morphology grid
  shows the combination of classes (Disk, Spheroid, Asymmetric,
  Irregular, Point Source) chosen by each of the 21 classifiers.  The
  histograms then show the morphology and interaction classes
  described in \S4 (recall that all sources identified as irregular
  are also considered to be asymmetric). The final consensus
  classifications (morphology = Irregular/Asymmetric, interaction =
  Interacting/Merging) are given by cyan stars.}
\end{figure}

For those sources classified as interacting or merging, we then
separated them into two additional subclasses: interacting/merging and
disturbed (interactions/mergers where the asymmetric, double nuclei,
or tidal arm flags had been selected as well) and interacting/merging
yet relatively undisturbed (interactions/mergers where none of the
asymmetric, double nuclei, or tidal arm flags were selected).  In
general, the latter class tends to catch early and/or wide separation
mergers as well as minor mergers.

We plot in Figure 4 the WFC3 F160W thumbnail images of our AGN sample,
broken down by interaction type, and also identify those AGN that are
X-ray and/or infrared selected.

\begin{figure*}[ht!]
\centering 
\gridline{\fig{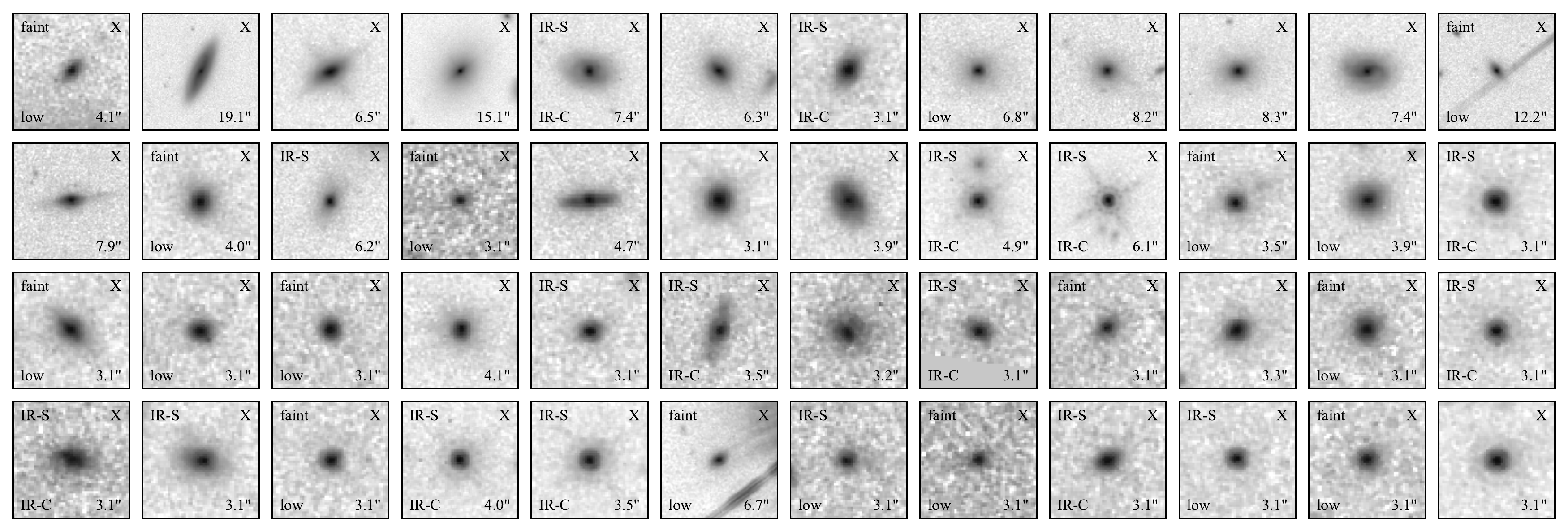}{\textwidth}{(a) Undisturbed}}
\gridline{\fig{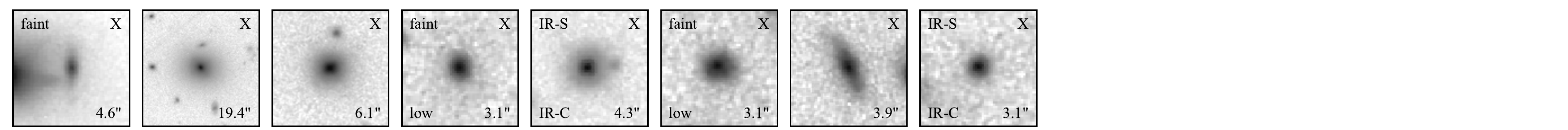}{\textwidth}{(b) Undisturbed 
  with a Companion (some of which fall outside this field of view)}}
\gridline{\fig{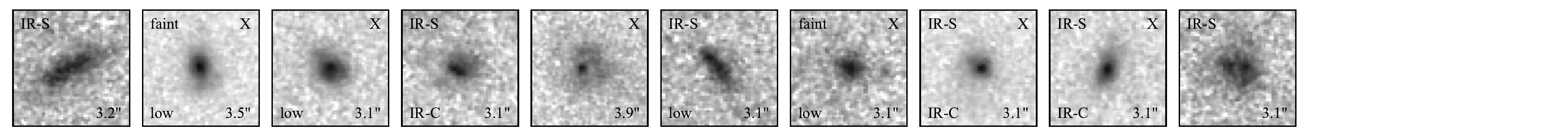}{\textwidth}{(c) Disturbed}}
\gridline{\fig{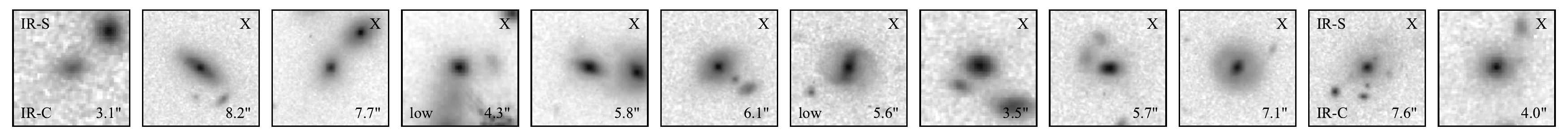}{\textwidth}{(d) Interacting/Merging yet Relatively Undisturbed}}
\gridline{\fig{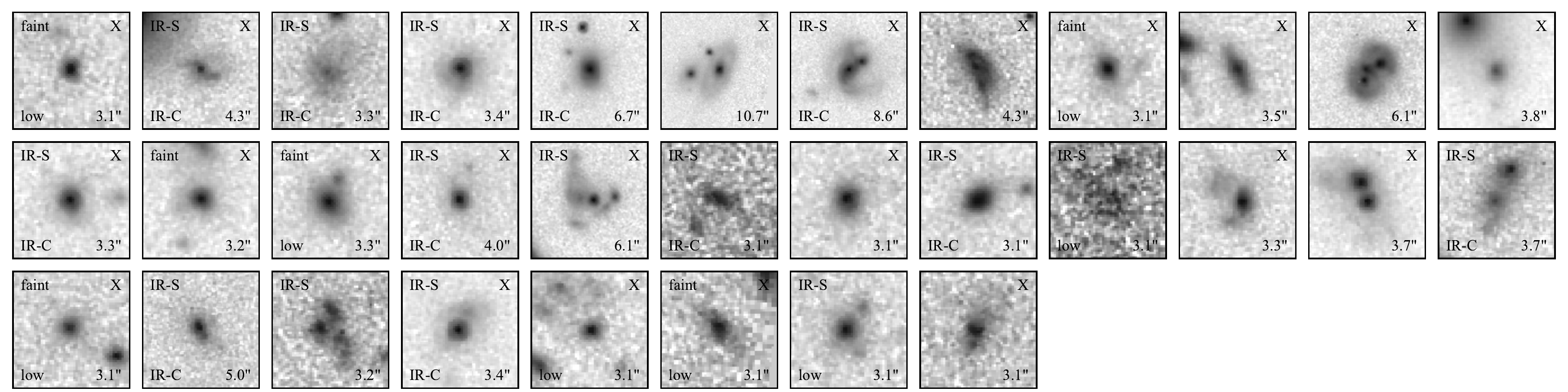}{\textwidth}{(e) Interacting/Merging and Disturbed}}
\caption{WFC3 F160W thumbnail images for the X-ray and IR-selected AGN
  samples, broken down by the consensus interaction class. Sources
  detected in the X-ray have an `X' in the upper right corner.  IRAGN
  selected using \cite{sanders07} are indicated with an `IR-S' in the
  upper left, where `faint' means that the source falls below the IRAC
  detection limits in at least one band.  IRAGN also selected using
  COSMOS15 are indicated with an `IR-C' in the bottom left, where
  `low' means that the source has $S/N < 3$ in at least one band. The
  diameter of each image is given in the lower right.  Each thumbnail
  is scaled to the size of the galaxy with a minimum size of
  3.1\arcsec\ (see \cite{kartaltepe15} for more details).  It is worth
  noting that classifiers were shown thumbnails that match those shown
  here, as well as an H-band thumbnail twice as large to help identify
  potential companions.}
\end{figure*}

\subsection{Morphology by AGN Type}

We plot in Figure 5 two comparisons between the morphology and 
interaction classes for the X-ray and IR-selected AGN samples, where 
we separate the sample into IRAGN-only (no X-ray), X-ray+IRAGN, and 
X-ray only.  The fractions for each category and subsample are given 
in Table 1. The figure on the left compares our full X-ray and 
\cite{sanders07}--selected IRAGN samples with one exception: we remove 
any AGN with a spectroscopic Type 1 identification to address the 
potential issue of rest-frame optical emission from the AGN masking 
the underlying morphology of the system.  Removing the broad-line AGN 
(BLAGN) primarily impacts the unobscured, high-luminosity population 
that is both X-ray and IR-selected (15 AGN are known BLAGN, and 11 of 
these are also IRAGN).  Of these 15 sources, 8 were classified as 
point sources by more than half of the classifiers, and 6 additional 
sources were classified as having a point source component by more 
than 20\% of the classifiers.  While it can occasionally be difficult 
for classifiers to distinguish between spheroids and point sources,
the additional confirmation of a Type 1 spectrum indicates that we may 
not be seeing the underlying host galaxy emission in these systems. 
This restriction lowers the relative point source fraction for the 
X-ray+IR AGN sample and subsequently raises the fractions for the 
remaining morphologies. 

The figure on the right further restricts the IRAGN samples to those 
sources also selected using the COSMOS15 catalog (see \S3.1).  In this 
plot, we also remove any X-ray--only AGN that fall below our flux cuts 
in one or more of the \cite{sanders07} IRAC bands or that have $S/N <
3$ in one or more of the COSMOS15 IRAC bands.  

\newpage
\clearpage

\begin{centering}
\begin{deluxetable*}{lcccccc}
\tabletypesize{\footnotesize}
\tablewidth{0pt}
\tablecaption{Percentage of AGN that meet Morphology and Interaction Classes}
\tablehead{
\colhead{Class}      &
\colhead{\cite{sanders07}}       &
\colhead{\cite{sanders07}}       &
\colhead{X-ray--only}             &
\colhead{COSMOS15}               &
\colhead{COSMOS15}               &
\colhead{X-ray--only}            \\
\colhead{}                                &
\colhead{IR-only AGN}            &
\colhead{IR+X-ray}                  &
\colhead{AGN}                         & 
\colhead{IR-only AGN}            &
\colhead{IR+X-ray}                  &
\colhead{AGN (restricted)\tablenotemark{1}}  \\
\colhead{}                                &
\colhead{}                                &
\colhead{(no BLAGN)}                                &
\colhead{(no BLAGN)}                                &
\colhead{}                                &
\colhead{no BLAGN}                                &
\colhead{no BLAGN}                               
}
\startdata
Number of Sources                          & 16         &   16     & 64     &     9   &   12   &   33 \\  
\tableline                                      
Disk                                                 & 38$^{+13}_{-10}$ & 44$^{+12}_{-11}$ & 38$^{+6}_{-6}$ & 44$^{+16}_{-14}$ & 46$^{+13}_{-12}$ & 45$^{+9}_{-8}$ \\
Spheroid                                          & 31$^{+13}_{-9}$ & 69$^{+9}_{-13}$ & 77$^{+4}_{-6}$ & 33$^{+18}_{-11}$ & 69$^{+10}_{-15}$ & 79$^{+5}_{-9}$ \\
Irregular                                          & 50$^{+12}_{-12}$ & 12$^{+13}_{-4}$ & 9$^{+5}_{-2}$ & 56$^{+14}_{-16}$ & 8$^{+14}_{-3}$ & 18$^{+9}_{-5}$ \\
Point Source                                    & 6$^{+12}_{-2}$ & 6$^{+12}_{-2}$ & 2$^{+3}_{-0}$ & 11$^{+18}_{-4}$ & 8$^{+14}_{-3}$ & 0$^{+5}_{-1}$ \\
Asymmetric                                     & 69$^{+9}_{-13}$ & 25$^{+13}_{-8}$ & 17$^{+6}_{-4}$ & 56$^{+14}_{-16}$ & 23$^{+15}_{-8}$ & 21$^{+9}_{-5}$ \\
Unclassifiable                                  & 0$^{+10}_{-1}$ & 0$^{+10}_{-1}$ & 2$^{+3}_{-0.0}$ & 0$^{+17}_{-2}$ & 0$^{+12}_{-1}$ & 0$^{+5}_{-1}$ \\
\tableline
Undisturbed                                     & 19$^{+13}_{-6}$ & 44$^{+12}_{-11}$ & 45$^{+6}_{-6}$ & 22$^{+18}_{-8}$ & 38$^{+14}_{-11}$ & 39$^{+9}_{-8}$ \\
Undisturbed + Companion               & 0$^{+10}_{-1}$ & 6$^{+12}_{-2}$ & 8$^{+5}_{-2}$ & 0$^{+17}_{-2}$ & 8$^{+14}_{-3}$ & 9$^{+8}_{-3}$ \\
Disturbed (D)                                    & 25$^{+13}_{-8}$ & 6$^{+12}_{-2}$ & 6$^{+5}_{-2}$ & 11$^{+18}_{-4}$ & 8$^{+14}_{-3}$ & 3$^{+6}_{-1}$ \\
Interacting/Merging                         & 56$^{+11}_{-12}$ & 44$^{+12}_{-11}$ & 39$^{+6}_{-6}$ & 67$^{+11}_{-18}$ & 46$^{+13}_{-12}$ & 48$^{+9}_{-8}$ \\
Interacting/Merging, Undisturbed    & 6$^{+12}_{-2}$ & 0$^{+10}_{-1}$ & 14$^{+5}_{-3}$ & 11$^{+18}_{-4}$ & 0$^{+12}_{-1}$ & 21$^{+9}_{-5}$ \\
Interacting/Merging and Disturbed (IMD) & 50$^{+12}_{-12}$ & 44$^{+12}_{-11}$ & 25$^{+6}_{-5}$ & 56$^{+14}_{-16}$ & 46$^{+13}_{-12}$ & 27$^{+9}_{-6}$ \\
IMD or D                                            & 75$^{+8}_{-13}$  &  50$^{+12}_{-12}$  & 31$^{+6}_{-6}$ & 67$^{+11}_{-18}$ & 54$^{+12}_{-13}$ & 30$^{+9}_{-7}$
\enddata
\tablenotetext{1}{The X-ray AGN in this column have been restricted to those sources
  that meet the \cite{sanders07} IRAC flux cuts and the COSMOS15 S/N cuts.}
\end{deluxetable*}
\end{centering}

\noindent Because these X-ray AGN are faint in the IR, we cannot
determine whether or not they would meet the IRAGN criteria.  This
restriction impacts only the X-ray--only sample and results in subtle
changes to the morphology and interaction classes.  In fact, we placed
this restriction on the X-ray sources primary to illustrate the fact
that it has little impact on the results, aside from increasing the
binomial confidence error bars calculated using the method of
\cite{cameron11}.

\begin{figure*}[t!]
$\begin{array}{cc}
\includegraphics[width=0.51\linewidth]{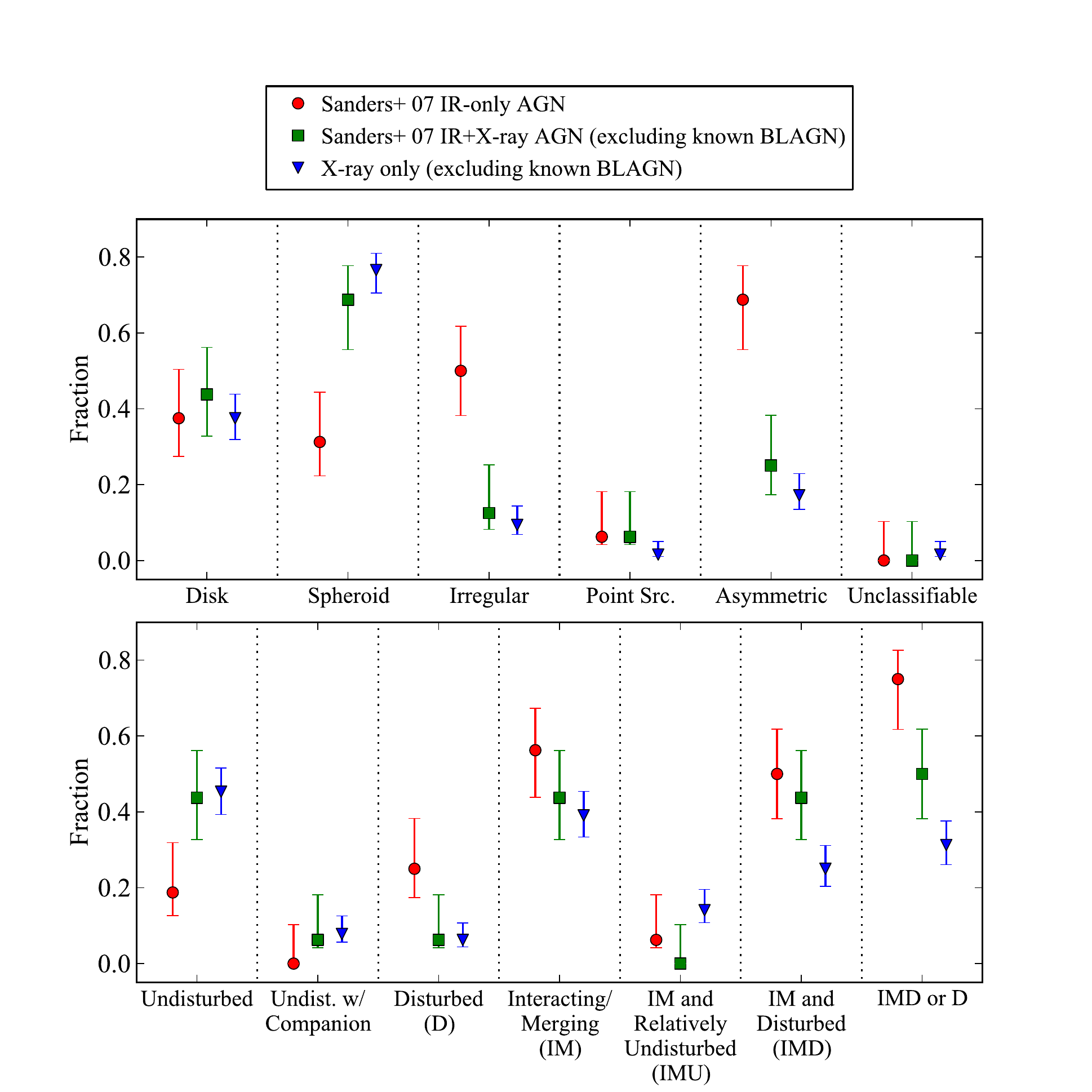} 
\includegraphics[width=0.51\linewidth]{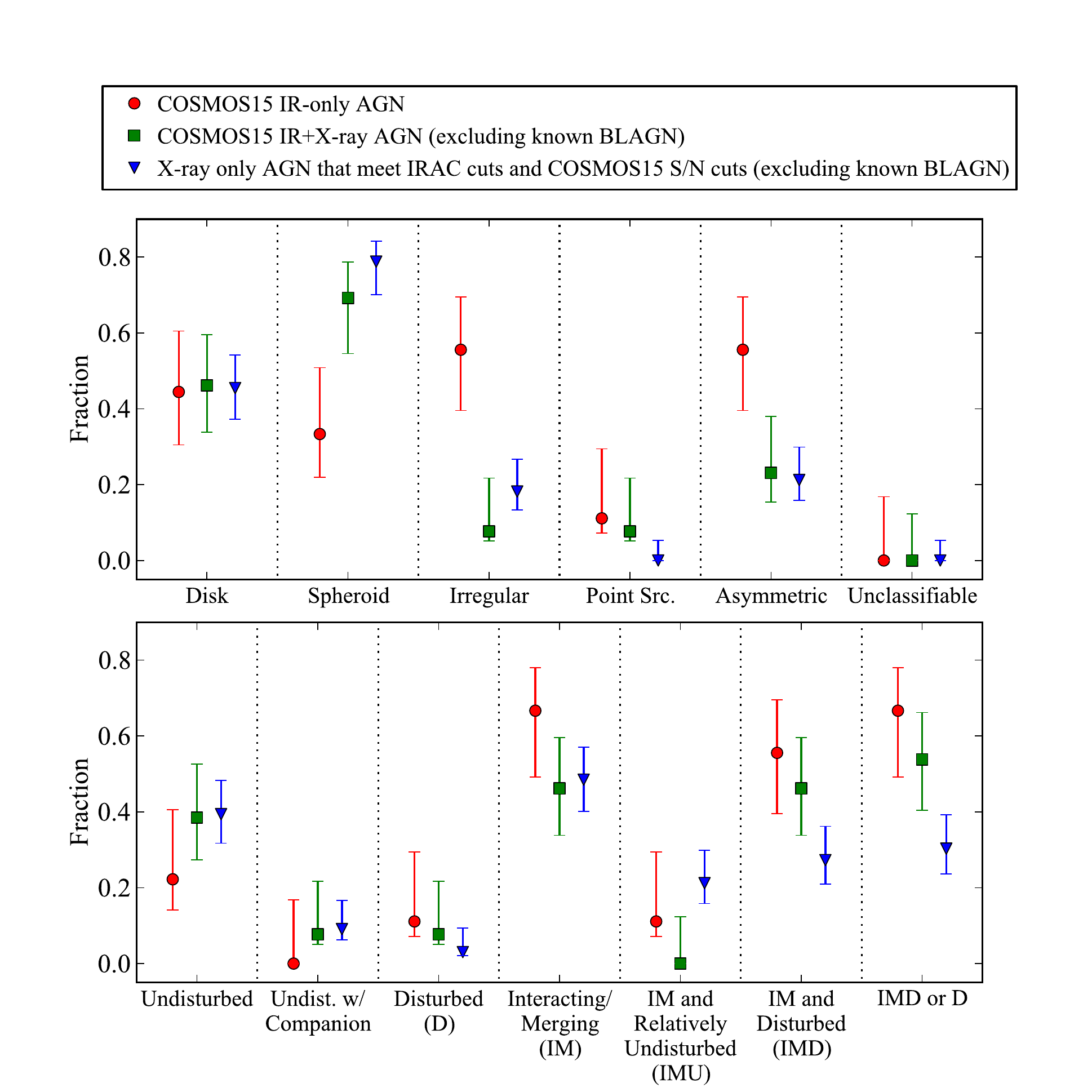} 
\end{array}$
\caption{Comparison between the fraction of AGN in the various
  morphology (disk, spheroid, irregular, point source, asymmetric, and
  unclassifiable) and interaction (undisturbed, undisturbed with a
  companion, disturbed, interacting/merging,
  interacting/merging+relatively undisturbed, and interacting/merging
  and disturbed) classes, broken down by AGN sample (IR-only,
  IR+X-ray, X-ray--only).  The plot on the left shows our full X-ray
  and \cite{sanders07}--selected IRAGN samples (with the exception of
  known BLAGN), whereas the plot on the right further restricts the
  IRAGN sample to those sources also identified using the COSMOS15
  catalog and removes any X-ray AGN whose \cite{sanders07} IRAC fluxes
  or COSMOS15 $S/N$ are too low to determine if they would meet the
  IRAGN selection criteria.  Binomial confidence error bars are
  calculated using the method of \cite{cameron11}.}
\end{figure*}

As can be seen from Figure 5, the main conclusions of this study are 
insensitive to whether the IRAGN are selected from \cite{sanders07} or 
COSMOS15.  There are of course subtle differences between the two 
samples, and the lower sample size for the COSMOS15 IRAGN increases 
the error bars, but the trends discussed below hold regardless of the 
catalog we use to identify IRAGN. 

Focusing first on the morphological classes, we see that the IR--only
AGN, which tend to be heavily obscured, high-luminosity, and
high-redshift AGN, are significantly more likely than X-ray--only AGN
to have been classified as irregular (50$^{+12}_{-12}$\%
vs. 9$^{+5}_{-2}$\%) or asymmetric (69$^{+9}_{-13}$\%
vs. 17$^{+6}_{-4}$\%), and are significantly less likely to have been
classified as having a spheroidal component (31$^{+13}_{-9}$\%
vs. 77$^{+4}_{-6}$\%) (all at the $>3\sigma$ level).  Their disk
fraction is indistinguishable from the other samples.  As for their
interaction class, these luminous, heavily-obscured AGN are less
likely than X-ray--only AGN to be undisturbed (19$^{+13}_{-6}$\%
vs. 45$^{+6}_{-6}$\%) and are more likely to be both `disturbed'
(25$^{+13}_{-8}$\% vs. 6$^{+5}_{-2}$\%), and `interacting/merging and
disturbed' (50$^{+12}_{-12}$\% vs. 25$^{+6}_{-5}$\%), though these
differences are only significant at the $\sim 2\sigma$ level.
However, if we combine those AGN classified as either `disturbed',
which may represent the late phases of a major merger, and
`interacting/merging and disturbed', we find that 75$^{+8}_{-13}$\% of
the IR-only AGN show signs of disturbance compared to only
31$^{+6}_{-6}$\% of the X-ray--only sample, a difference that is
significant at the $3\sigma$ level. The vast majority of our
admittedly small sample of IR-only AGN therefore show signs of clear
merger activity and/or disturbances that may be indicative of recent
mergers.  This indicates that major mergers may indeed play a dominant
role in fueling high-luminosity, heavily-obscured AGN activity.

The morphologies and interactions classes of the AGN that meet both
the X-ray and IRAGN criteria tend to fall between those of the
X-ray--only and IR--only samples.  This implies that while
Seyfert-luminosity AGN are not predominantly associated with
interacting and/or heavily disturbed hosts, the fraction of AGN with
disturbed morphologies may increase at higher luminosities/redshifts
(i.e. the X-ray+IR sample) or as the nuclear obscuration increases
(IRAGN--only).  We examine the independent impacts of luminosity and
obscuration in \S4.2.

Finally, it is worth noting that the increased merger fraction for our
luminous, heavily obscured IR-only sample remains if we focus only on
AGN in the fixed redshift range of $z=1.5-2.5$ where a majority of
IR-only AGN lie.  Of the 8 IR-only AGN in this redshift range (6 of
which are selected both from the \cite{sanders07} and COSMOS15
catalogs), 6 (75\%) are interacting or merging and disturbed. In
contrast, only 2/12 X-ray+IR AGN (17\%) are interacting (one is
disturbed and the other is relatively undisturbed), and only 4/17
(24\%) of the X-ray--only AGN are interacting (3 are disturbed, and
one is relatively undisturbed).  In this redshift range, obscuration
(e.g., X-ray detected or not) appears to play a larger role than
luminosity (e.g., IRAGN or not), though these results may be biased by
the small number of sources.  Nevertheless, the significantly higher
merger fraction among luminous and obscured IR-only AGN in this
limited redshift range suggests that the results for our full sample
have not been highly biased by the larger average redshift of this
sample compared to that of the X-ray selected AGN population.

\begin{figure*}[t!]
$\begin{array}{cc}
\includegraphics[width=0.5\linewidth]{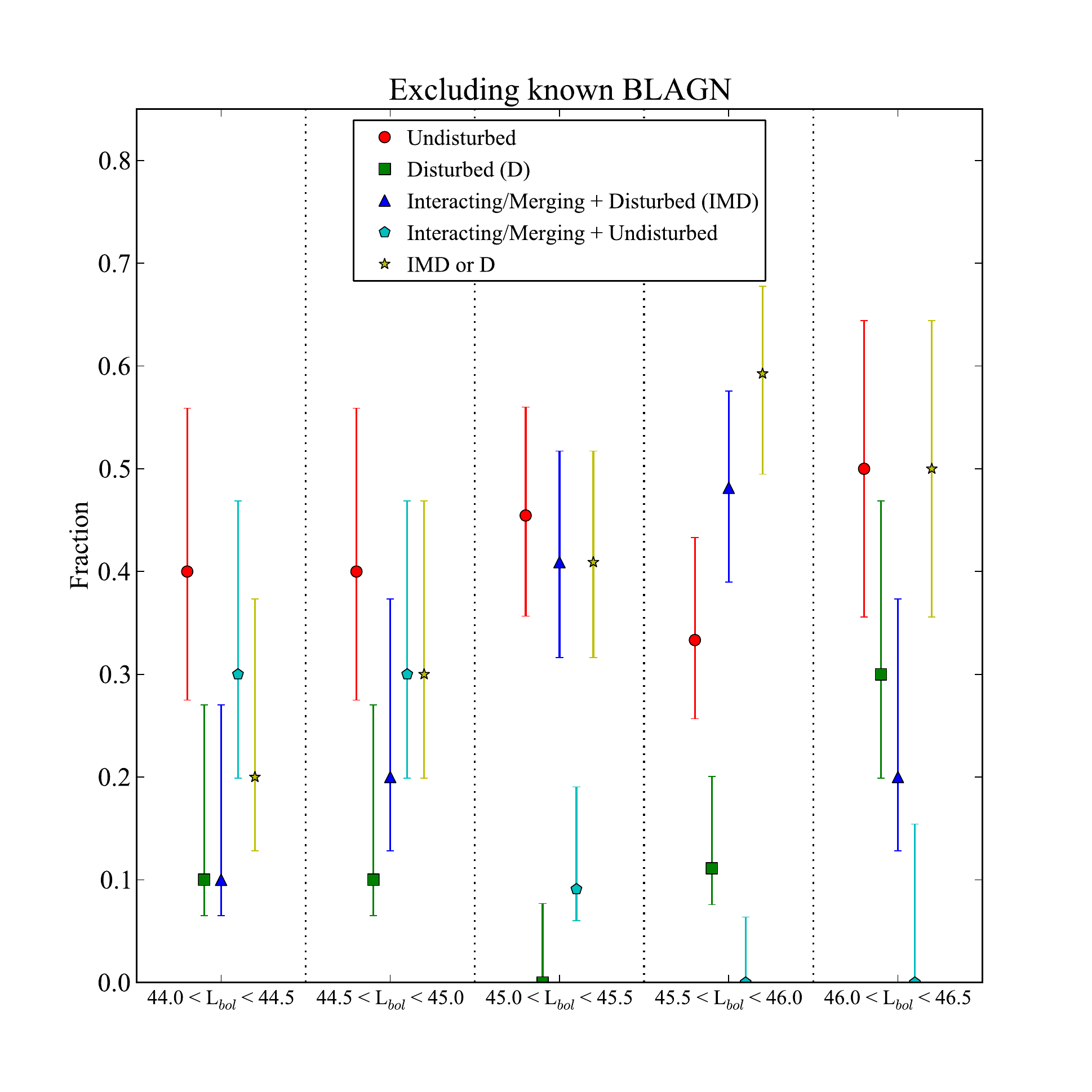} 
\includegraphics[width=0.5\linewidth]{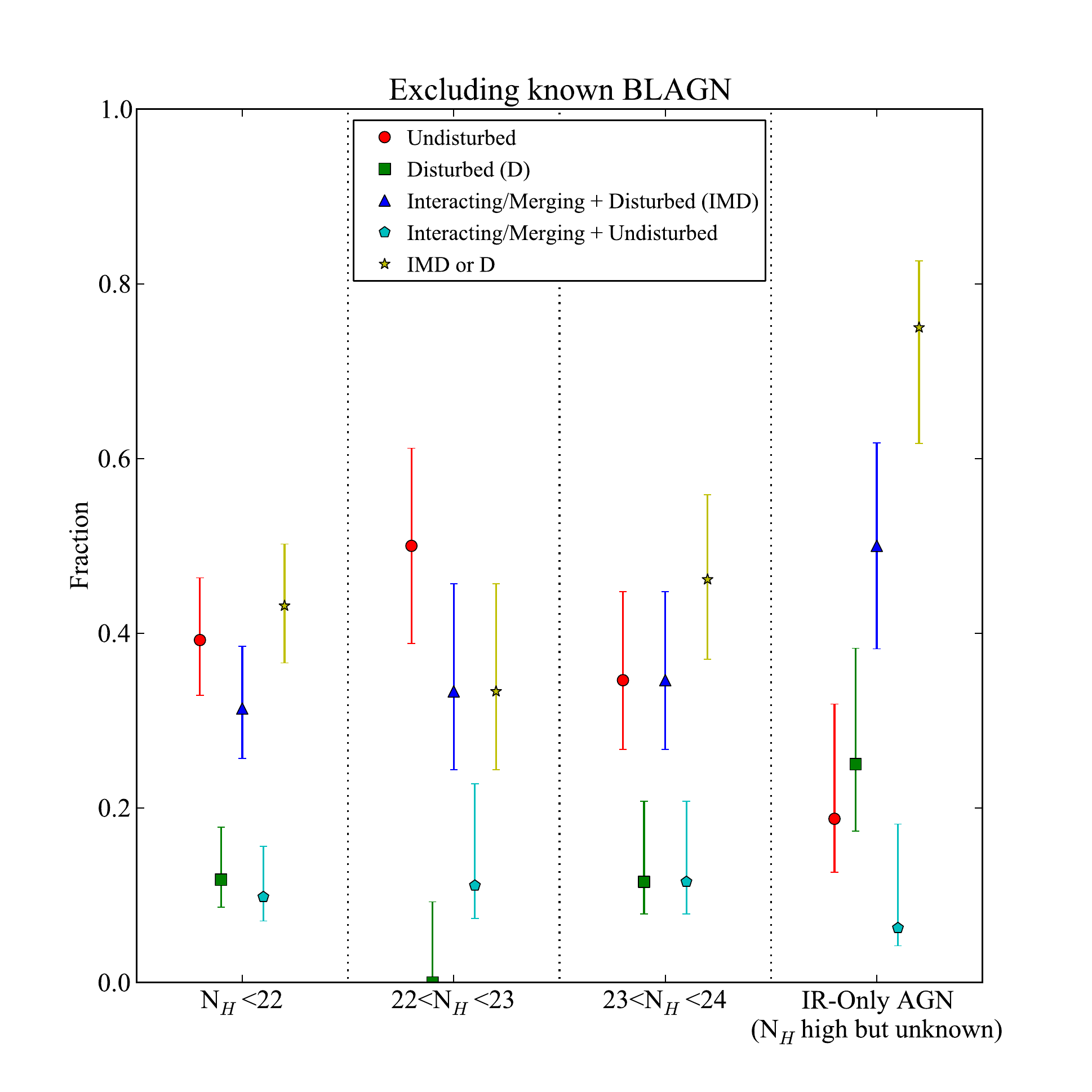} 
\end{array}$
\caption{AGN interaction classes as a function of AGN bolometric
  luminosity in units of ergs~s$^{-1}$ (left) and AGN obscuration in
  units of log cm$^{-2}$ (right). As in Figure 5, we exclude known
  BLAGN from the sample (see \S4.1). While both luminosity and
  obscuration may impact the morphology of our sample, luminosity and
  obscuration are not strictly independent in our sample due to the
  inclusion of the IR-only AGN (high luminosity/high obscuration AGN).  A
  larger and less biased sample is required to isolate the
  effects of luminosity and obscuration. }
\vspace{0.5cm}
\end{figure*}

\subsection{Effects of Luminosity and Obscuration on Morphology}

To determine if we can isolate the effects of luminosity and
obscuration on the differences between the IR-only (high luminosity,
heavily obscured) and X-ray only (lower luminosity, less obscured)
populations, we plot in Figure 6 four of the the interaction classes
as a function of both AGN bolometric luminosity (for luminosity bins
with at least 5 AGN) and obscuration.  We also plot the sum of AGN
classified as either `interacting/merging and disturbed' or simply
`disturbed'.  For consistency with Figure 5, we exclude known BLAGN
(see also \S4.1).

The fraction of AGN classified as undisturbed appears to be
independent of an AGN's bolometric luminosity, and while the fraction
that are interacting/merging \textit{and} disturbed (IMD) increases
with luminosity, this drops again in the highest luminosity bin.  This
drop is offset in part by a rise in AGN classified simply as disturbed
(D) in the highest luminosity bin, such that the combination of
`interacting/merging and disturbed' plus `disturbed' remains high at
the highest luminosities.  This trend is the opposite of those AGN
classified as `interacting/merging yet relatively undisturbed', which
appear to be predominantly lower-luminosity Seyfert galaxies.
However, luminosity and obscuration are not strictly independent in
our sample: heavily obscured IR-only AGN comprise a significant
fraction of the two highest luminosity bins, and our sample does not
contain lower-luminosity AGN too obscured to be detected in the X-ray.
It is therefore possible that the apparent increase in the disturbed
fraction (IMD or D) at high luminosity is due at least in part to the
heavily obscured IR-only AGN in our sample.  Indeed, the rise in IMD+D
with luminosity is still present but not as pronounced if we consider
only the X-ray selected AGN.

Quantifying obscuration is somewhat more difficult than quantifying
luminosity.  \cite{marchesi16b} estimate X-ray column densities via
X-ray spectral fitting for sources with $>30$ counts in the 0.5-7~keV
band, but only 55\% of our X-ray sources meet this criterion.  For the
remaining X-ray sources, we adopt the column density estimates from
\cite{marchesi16a} which are based on the observed X-ray hardness
ratios (or limits on the hardness ratios).  If only an upper limit is
available (9\% of the X-ray sample), we take this to be consistent
with no obscuration, and in cases where only a lower limit is
available (13\% of the X-ray sample), we adopt this lower limit as our
measure of N$_{\rm H}$.  Furthermore, while we expect that the X-ray
non-detected IRAGN are heavily obscured, we do not know precisely how
obscured they are. Given these limitations, we plot in Figure 6b the
samples with N$_{\rm H} < 10^{22}$, $10^{22} < $N$_{\rm H} < 10^{23}$,
$10^{23} < $N$_{\rm H} < 10^{24}$, along with the IR-only AGN.

While the undisturbed fraction is lowest for the heavily obscured
IR-only AGN, this trend is not statistically significant.  Likewise,
the `disturbed' and `interacting/merging and disturbed' fractions are
highest for the IR-only population, but only marginally so, and while
the sum of these disturbed categories (IMD+D) is highest for the
IR-only AGN, there is no clear trend with obscuration for the X-ray
detected AGN.

Disentangling the effects of luminosity and obscuration is therefore
challenging, both due to our small sample size and the correlation
between luminosity and obscuration, particularly for the IR-only
subsample.  It appears plausible that both luminosity and obscuration
impact the morphology of our sample, but far more complete samples
lacking a bias against low-luminosity, heavily obscured AGN would be
required to draw a definitive conclusion (see, for instance,
\cite{kocevski15}, who find that more heavily obscured AGN show an
increase in disturbed morphologies.)

\subsection{Comparison to Previous Results}

\begin{figure*}[t!]
$\begin{array}{cc}
\includegraphics[width=0.5\linewidth]{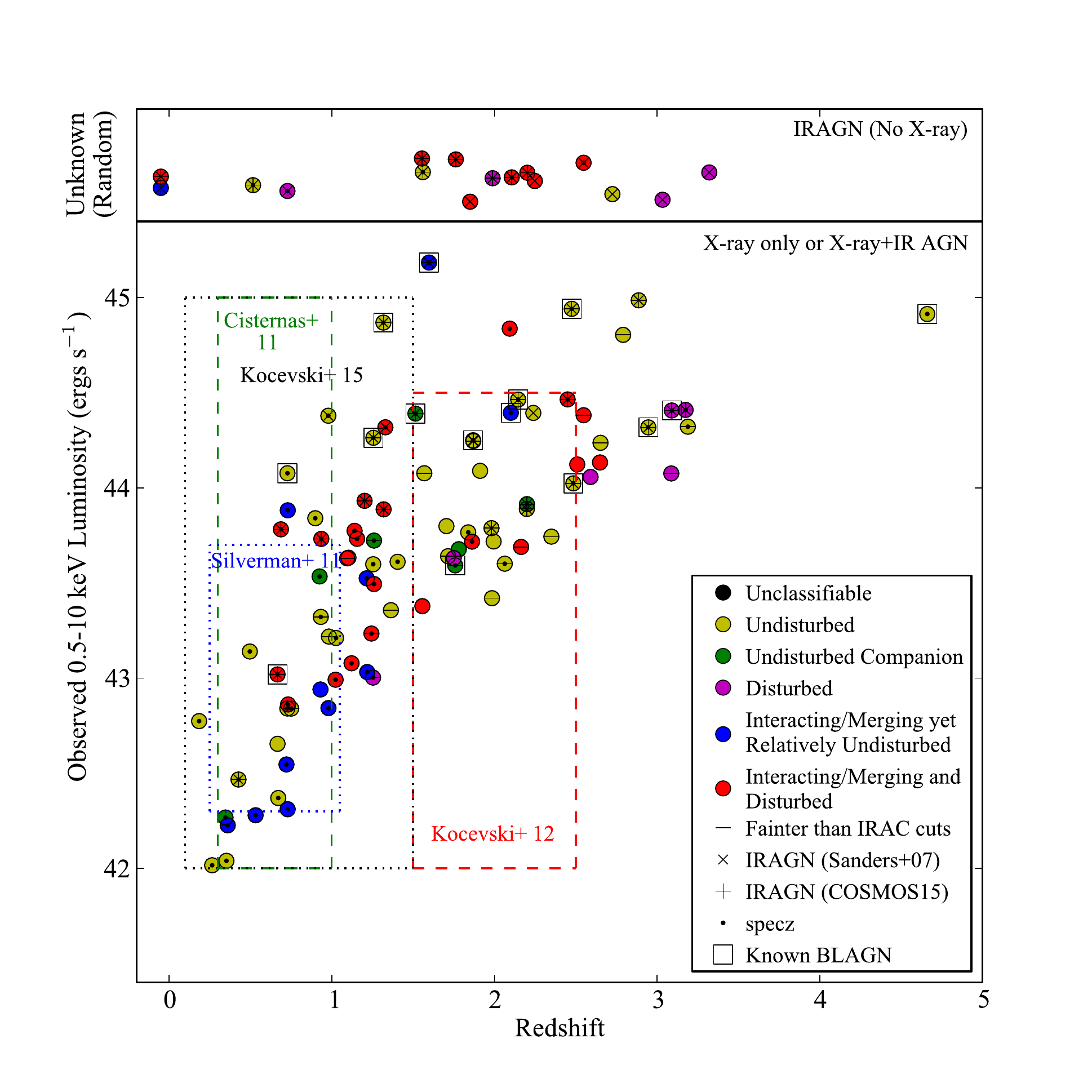} 
\includegraphics[width=0.5\linewidth]{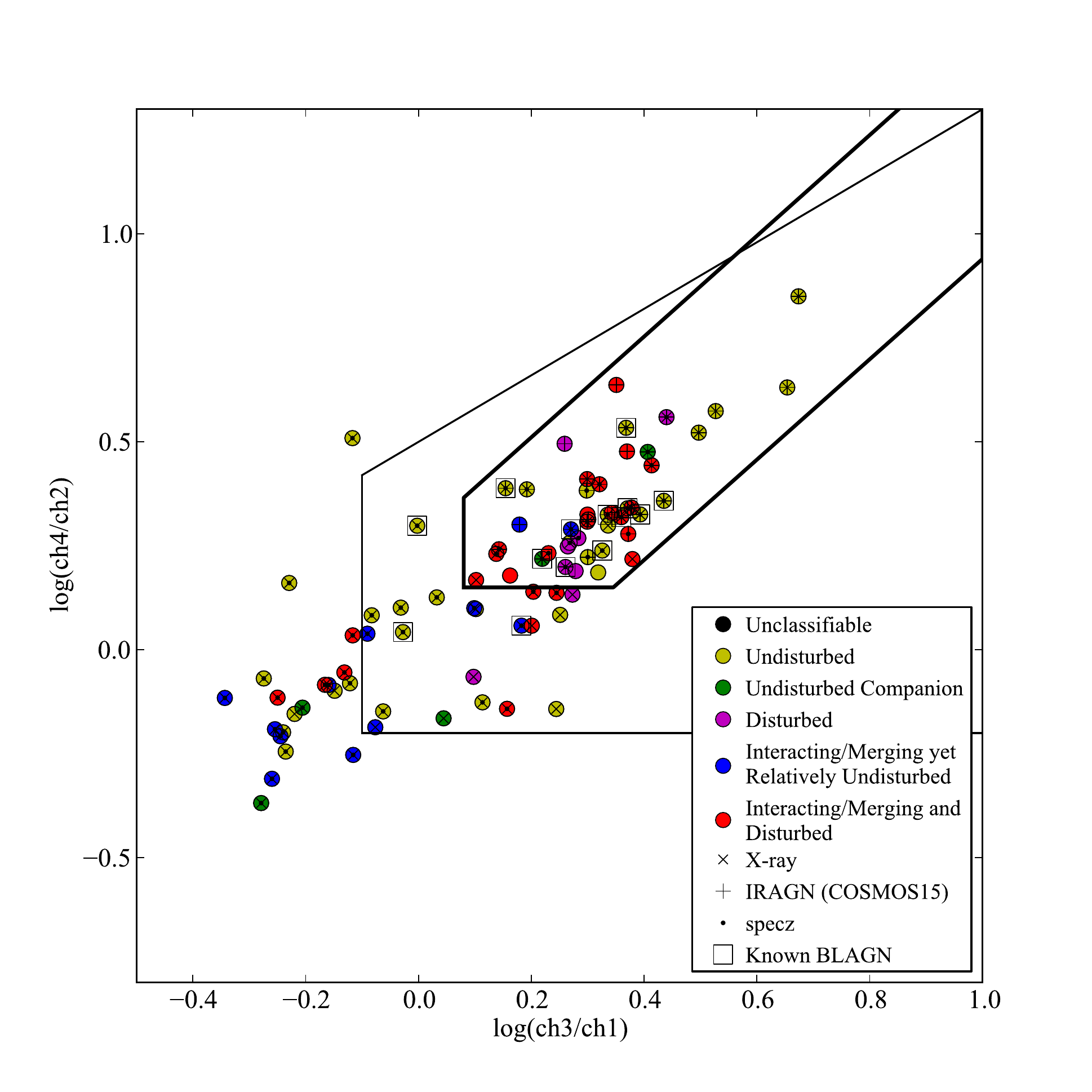} 
\end{array}$
\caption{Consensus interaction class for our full sample as a function
  of both X-ray luminosity and redshift (left) and IRAC color
  (right). The regimes samples by several previous studies are shown
  as dotted lines.  In addition to interaction class, we indicate
  those AGN that are X-ray or IR-detected, as well as those with
  spectroscopic redshifts and those known to be BLAGN. The AGN
  selection region from \cite{donley12} is plotted on the IRAC color
  plot on the right, as is the larger AGN wedge of \cite{lacy07}.}
\vspace{0.5cm}
\end{figure*}

We plot in Figure 7 the consensus morphology classes of our sample as
a function of redshift/X-ray luminosity and IRAC color, and overplot
the redshift and luminosity regimes sampled by the
\cite{cisternas11b}, \cite{silverman11}, \cite{kocevski12}, and
\cite{kocevski15} studies.  We see good agreement when we compare our
morphology assessments to these prior studies.  \cite{silverman11}
found that $18\% \pm 8\%$ of $0.25<z<1.05$ Seyferts are in kinematic
pairs (early mergers).  Of the 16 AGN in our sample that meet their
selection criteria, 7 (44\%) are interacting. However, 2 of these are
in late mergers and 1 is in a minor merger, for an early merger
fraction of 4/16, or 25\%, consistent with the \cite{silverman11}
result.

In a similar redshift regime that extends to both higher and lower
luminosities, \cite{cisternas11b} found that 54\% of AGN were
undisturbed, 31\% were mildly distorted, and 15\% were strongly
distorted.  Of the 17 sources in our sample that meet their selection
criteria, 9 (53\%) are undisturbed, 3 (18\%) are interacting/merging
yet relatively undisturbed, and 5 (29\%) are interacting/merging and
disturbed.  However, 2 of these interacting/merging and disturbed AGN
are relatively minor disturbances that may have fallen in the 'mildly
distorted' bin.  Generally, we are again in good agreement with this
prior study. 

A direct comparison to the \cite{kocevski15} study of $z<1.5$ X-ray
selected AGN is difficult as their sample is split into subsamples
with different X-ray obscurations, with each subsample covering a
broad range of both luminosities and redshifts.  If we focus only on
their most obscured AGN with N$_{\rm H} > 23.5$~cm$^{-2}$ and compare
to our IR--only sample that falls within the same redshift bounds, we
see a larger merger fraction (50\%) than they report ($22\%$).
However, our heavily-obscured IR--only AGN are predominantly quasars,
whereas a significant fraction of their highly-obscured AGN have
Seyfert-like luminosities. The higher merger fraction observed in our
work may therefore be due to the systematically higher luminosity of
our sample (see \S1 and the references therein).

Finally, \cite{kocevski12} looked at the CANDELS morphologies of
higher redshift X-ray AGN.  They find that 19\% of $L_x >
10^{43}$~erg~s$^{-1}$ AGN are interacting/merging and 47\% are
undisturbed.  Of the 26 sources in our sample that meet their
selection criteria, 5 (19\%) are interacting/merging, and 16 (62\%)
are undisturbed.  We therefore conclude that our findings for lower
luminosity and/or lower redshift AGN are consistent with findings in
the literature that conclude that mergers do not play a dominant role
in the fueling of Seyfert galaxies at either low ($z<1$) or high
($z\sim2$) redshift, even if they may be responsible for driving AGN
activity in higher luminosity, higher redshift, and more heavily
obscured AGN.

\subsection{Undisturbed Disks}

Constraining the role of major mergers in fueling AGN activity can be
complicated by a potential time delay between the merger and the peak
of AGN activity.  However, as disks are expected to be disrupted or
destroyed by major mergers, the undisturbed disk fraction can be used
to place a constraint on the fraction of AGN that are unlikely to have
undergone a major merger, at least in the recent past.  Of the
X-ray--only AGN, 16\% are classified as undisturbed galaxies with a
disk component, as are a similar fraction (4/27, or 15\%) of X-ray+IR
AGN.  Of the X-ray non-detected IRAGN, however, only 1/16 (6\%) is
undisturbed with a disk component.  The fraction of relatively
undisturbed disks is therefore low across our sample. However, a
non-negligible fraction of both X-ray--only and X-ray+IR AGN lie in
undisturbed disks, indicating that both moderate and high luminosity
AGN activity can occur in the absence of any recent major
interaction. The undisturbed disk fraction is even lower, however, for
the IR-only AGN, suggesting that it may be less common for these AGN
to be triggered in isolation.

\section{Conclusions}

In summary, using the deep CANDELS NIR imaging in the CANDELS/COSMOS
field, we have compared the rest-frame visual morphologies of X-ray
and IR-selected AGN.  The X-ray--only AGN in our sample cover a range
of redshifts and tend to be Seyfert luminosity AGN with low to
moderate obscuration.  They are the least likely to be disturbed and
the most likely to have a spheroidal component.  When they are
interacting or merging, the primary host galaxy often appears to
remain relatively undisturbed, either because the merger is
comparatively minor or because it is in an early phase more accurately
described as a close pair. This suggests that low obscuration Seyfert
luminosity activity either precedes the high-luminosity, heavily dust
enshrouded phase predicted during a major merger (for close pairs that
will later become a major merger), that low level AGN activity can be
triggered by minor interactions, or that this activity is unrelated to
the nearby companion.  Our findings for the X-ray sample are
consistent with past studies that have concluded that mergers are not
a dominant source of fueling at low AGN luminosity and obscuration
(e.g. \citealt{kocevski12,cisternas11b,silverman11}).

While stochastic fueling may account for Seyfert-luminosity AGN,
models of galaxy and AGN formation suggests that major mergers are the
dominant fueling mechanism for luminous, obscured AGN and their hosts
(e.g. \citealt{hopkins08}).  The IR-only AGN in our sample also span a
range of redshifts, but unlike the X-ray--only sample, they tend to be
high-luminosity, heavily obscured AGN.  These AGN are significantly
($>3\sigma$) more likely than X-ray--only AGN to have been classified
as irregular or asymmetric and are also more likely than X-ray--only
AGN (at the $\sim 2\sigma$ level) to be classified both as undergoing
interactions/mergers that significantly disrupt the host galaxy and as
simply `disturbed', which could potentially indicate the late stages
of a merger.  Combining these two categories, we find that 75\% of
IR-only AGN show significant signs of disturbance compared to only
31\% of the X-ray--only sample, a difference that is significant at
the $3\sigma$ level.  These results are consistent with theoretical
models of galaxy and AGN growth as well as with recent observational
evidence for an increase in the merger fraction at high luminosity
and/or obscuration \citep{guyon06, kartaltepe10, koss12,
  treister12,ellison16, fan16,koss10,
  urrutia12,satyapal14,kocevski15,ellison16,shangguan16,weston17}.

The lack of evidence for merger-driven AGN growth in typical
Seyfert-luminosity, X-ray selected AGN
\citep{schawinski11,kocevski12,simmons12,sanchez04, grogin05,
  pierce07,gabor09,cisternas11b,simmons12,villforth14, rosario15,
  bruce16} can therefore be attributed to looking for mergers among
the wrong population of AGN/hosts.  By targeting luminous and heavily
obscured AGN using IR selection, we have selected exactly the sample
of AGN most likely to be merger driven, and indeed find evidence that
the vast majority are heavily disturbed.

\acknowledgements

JLD acknowledges support for this work provided by NASA through Hubble
Fellowship grant HF-51303.01 awarded by the Space Telescope Science
Institute, which is operated by the Association of Universities for
Research in Astronomy, Inc., for NASA, under contract NAS5-26555.  BDS
acknowledges support from the National Aeronautics and Space
Administration (NASA) through Einstein Postdoctoral Fellowship Award
Number PF5-160143 issued by the Chandra X-ray Observatory Center,
which is operated by the Smithsonian Astrophysical Observatory for and
on behalf of NASA under contract NAS8-03060.  CL acknowledges support
from NSF grant AST-1004583.  Finally, we thank the anonymous referee
for discussions and comments that improved the paper.

\newpage

\bibliographystyle{apj}


\end{document}